\documentclass[3p,times,procedia]{elsarticle}
\usepackage{nupha_ecrc}
\usepackage{subcaption} 

\volume{00}

\firstpage{1}

\journalname{Nuclear Physics A}

\runauth{L. Apolin\'{a}rio}

\jid{nupha}

\jnltitlelogo{Nuclear Physics A}

\usepackage{amssymb}

\usepackage[figuresright]{rotating}

\begin{document}

\begin{frontmatter}



\dochead{XXVIIIth International Conference on Ultrarelativistic Nucleus-Nucleus Collisions\\ (Quark Matter 2019)}

\title{Road map to extracting medium properties: an overview}


\author[a,b]{Liliana Apolin\'{a}rio}
\address[a]{LIP, Av. Prof. Gama Pinto, 2, P-1649-003 Lisboa, Portugal}
\address[b]{Instituto Superior T\'{e}cnico (IST), Universidade de Lisboa, Avenida Rovisco Pais 1, 1049-001 Lisbon, Portugal}

\begin{abstract}
In this manuscript, it is presented an overview of the Quark-Gluon Plasma properties measured, so far, using hard probes. We will focus on both quantitative and qualitative properties that have been (or are about to be) measured, making a link between the theoretical description and experimental results. Throughout the manuscript, highlights to some of the conferences' results will be given, but without an extensive overview. A personal opinion of the most important developments and critical problems that need more work in the future is presented in the end.
\end{abstract}


\end{frontmatter}


\section{Introduction: Road so far}
\label{sec:intro}

The first measurements done at RHIC and the LHC allowed to extract one of the QGP properties whose characterisation is only possible by using hard probes: the jet transport coefficient. Since then, there has been a long evolution from (1) single-particle energy loss description to in-medium parton shower, (2) medium-induced gluon radiation to include medium response, (3) light and heavy-quark description, among others. For more information, we refer the reader to~\cite{Luo,Tywoniuk,Cao}. Simultaneously, there has been an increase of novel observables: single-particle measurements, fully reconstructed jets, calibrated probes without the need of proton-proton collision results as a reference (e.g: boson-jet correlations), intra-jet observables, among others (see~\cite{Wang,Chen,Trzeciak} for more details). With such a plethora of experimental results and qualitative breakthroughs in our understanding of how high momentum objects are modified when traversing a hot and dense medium, is time to pose ourselves the question: \textit{What have we learned about the QGP?}, in particular, \textit{What are the properties that hard probes allowed us to establish?} 

In this manuscript, I will present a tentative summary of our quantitative and qualitative understanding of the QGP properties achieved through the continuous and simultaneous development of hard probes description and observables. At the same time, we will try to highlight the questions that remain open. This will be section~\ref{sec:2}. A personal and biased view of the next steps in the short and mid-term will be presented in section~\ref{sec:3}. A final summary can be found in section~\ref{sec:4}.

\section{QGP Properties}
\label{sec:2}

\subsection{Transport coefficient}

The first QGP property to be measured by hard probes was the transport coefficient, $\hat{q}$, which measures the transparency of the medium to the passage of a high momentum particle. From this parameter, one can determine the \textit{strenght} of jet-medium interaction, either by elastic or inelastic scattering processes, by measuring the amount of energy loss and/or transverse momentum broadening induced by these mechanisms. The former, present in models that account for medium response, is usually denoted by $\hat{e} = d\left\langle E \right\rangle/dt$, where $E$ is the energy of the interacting high momentum particle and $t$ the in-medium path-length. The later is characterized by $\hat{q} = d \left\langle \Delta p_T^2 \right\rangle/dt$, where $\Delta p_T$ is the transverse momentum broadening acquired by multiple interactions with the medium. When the medium is almost in local equilibrium, these two parameters can be related through the medium temperature~\cite{Moore:2004tg,Qin:2009gw}, $T$: $\hat{q} \propto T \hat{e}$. 

\begin{figure}[h]
\centering
 \begin{subfigure}[c]{0.45\textwidth}
 \includegraphics[width=\textwidth]{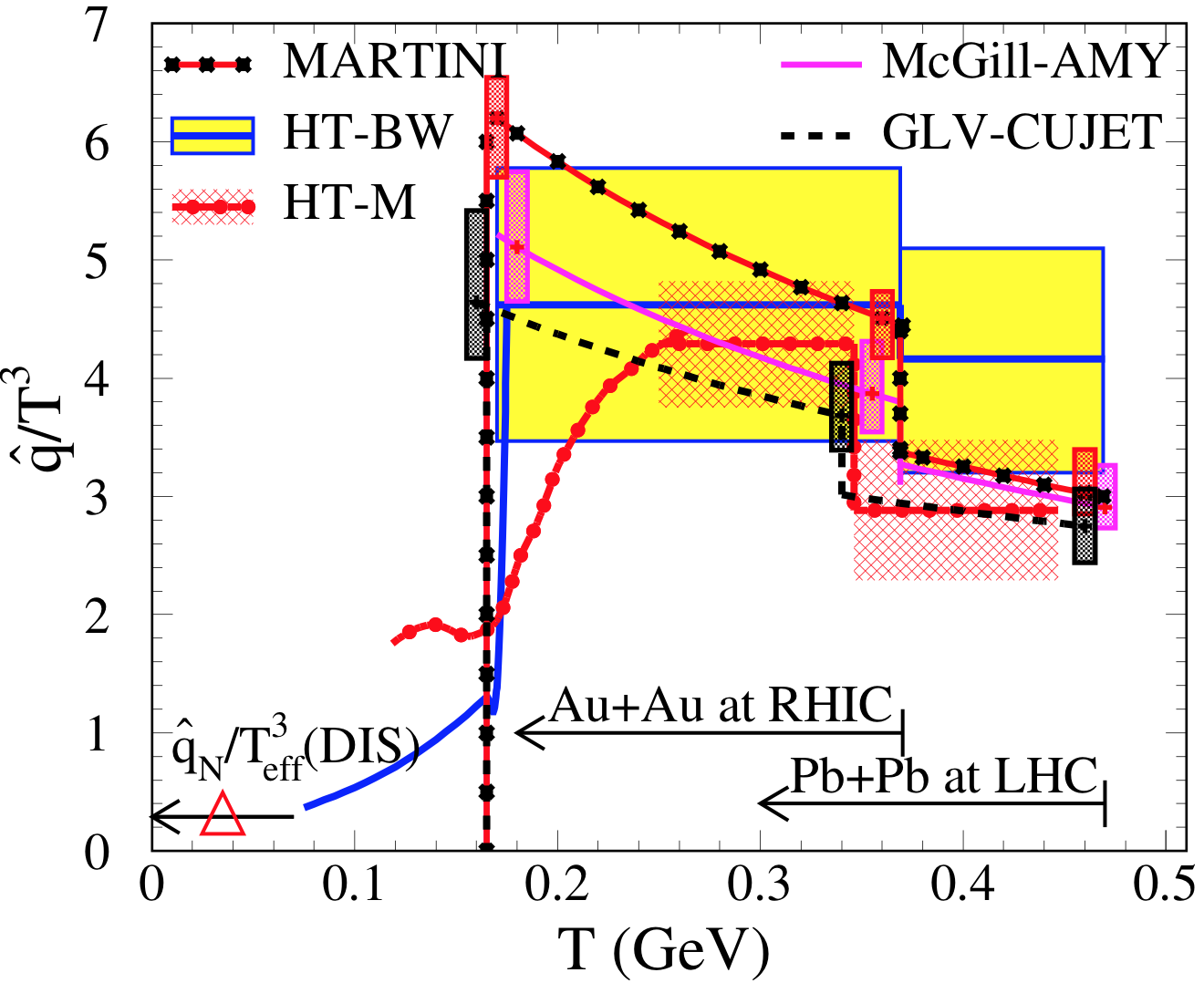}
 \caption{Temperature dependence of the scaled jet transport parameter $\hat{q}/T^3$ in different jet quenching models as provided by thet JET Collaboration~\cite{Burke:2013yra}. The arrows indicate the range of temperatures at the center of the most central nuclear collisions, for both RHIC and the LHC.\\}
 \label{fig:qhatJET}
 \end{subfigure}
 \hspace{5mm}
 \begin{subfigure}[c]{0.45\textwidth}
 \includegraphics[width=\textwidth]{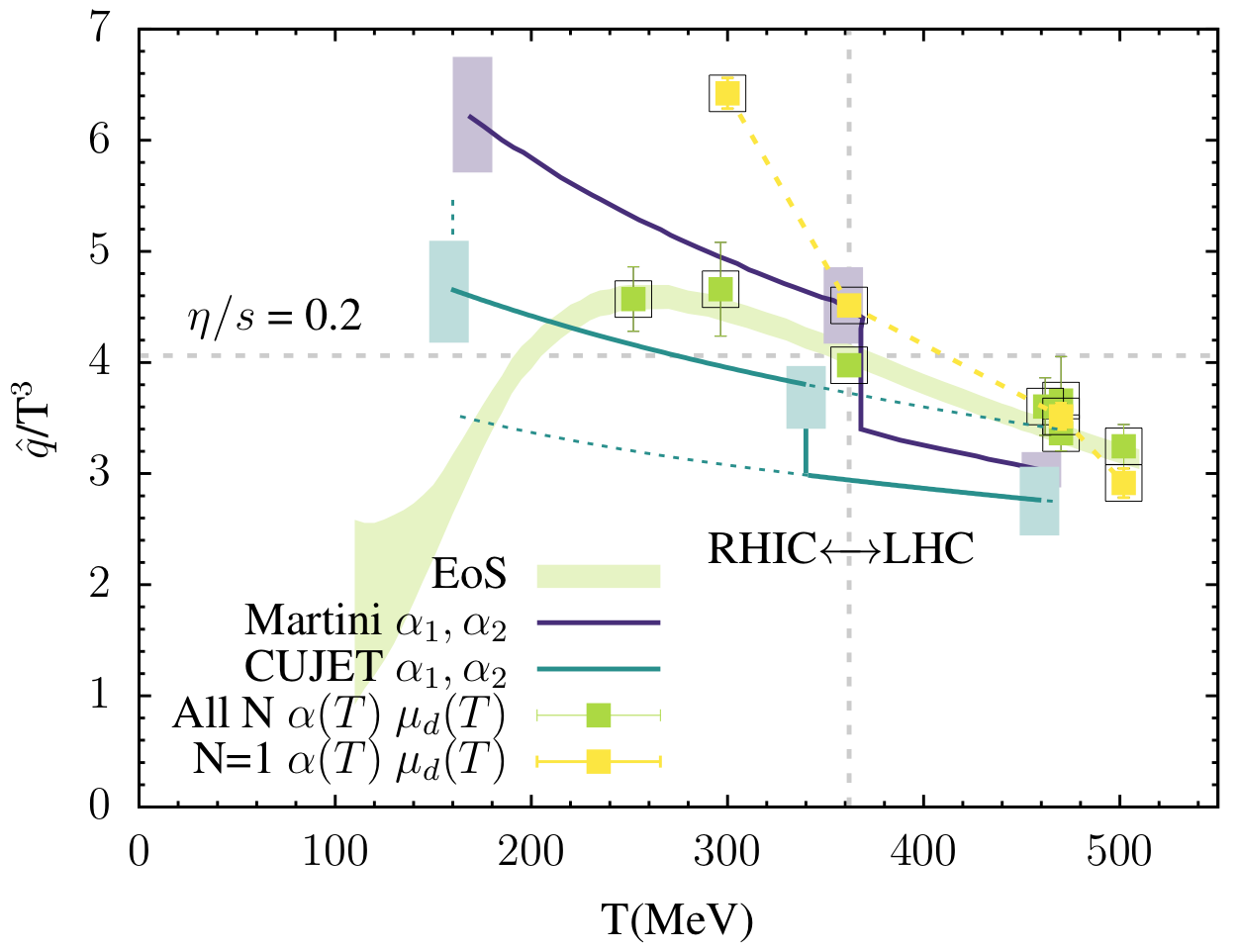}
 \caption{Temperature evolution of the QGP transport parameter $\hat{q}$ for a gluon with an energy of 10~GeV, when using the density extracted from an all order (green squares) or a fist order (yellow squares) analysis. 
 For comparison, the CUJET (blue) and MARTINI (purple) curves match the ones from Fig.~\ref{fig:qhatJET}. Fig. taken from~\cite{Feal:2019xfl}.\\}
 \label{fig:qhatFeal}
 \end{subfigure}
 \caption{Evolution of the scaled transport coefficient with the QGP temperature.}
 \label{fig:qhat}
\end{figure}

One of the first attempts to extract this parameter from single hadron spectra was performed by the JET collaboration~\cite{Burke:2013yra}. By using five semi-analytical approaches, coupled with hydrodynamic simulations, it was possible to extract the evolution of $\hat{q}$ as a function of the temperature, for both Au+Au collisions at RHIC and Pb+Pb collisions at the LHC. The result is shown in Fig.~\ref{fig:qhatJET}, where the average $\hat{q}_0$ was identified as $\hat{q}_0 \simeq 1.2 \pm 0.3$, at an initial temperature of $T_i = 370$~MeV (RHIC), and $\hat{q}_0 \simeq 1.9 \pm 0.9$, at an initial temperature of $T_i = 470$~MeV (LHC), both at an initial time of $\tau_0 = 0.6$~fm/c. Despite the accurate determination of this parameter (particularly when taking into account the different temperature evolutions provided by each approach), a puzzling observation was that the $\hat{q}$ parameter was showing a discontinuity when going from RHIC to LHC centre-of-mass energies. This would thus indicate that, by preserving the expected continuous evolution of $\hat{q}$ with temperature, it would lead to a higher $\hat{q}$ at RHIC, then at the LHC, when measured at the same temperature. This fact was later suggested in~\cite{Andres:2016iys}, thus evidencing the possibility of the centre-of-mass collision energy dependency of $\hat{q}$.

A common limitation of the analytical approaches where such extraction of the $\hat{q}$ parameter was based, is the approximation in the number of in-medium scattering centres. Two approximations were usually employed: a few collection of scattering centres, or, in the opposite limit, multiple in-medium soft scatterings (Gaussian approximation). Most recently, there were several attempts to re-sum all terms in opacity expansion beyond the Gaussian approximation. These include~\cite{Feal:2019xfl,Sievert:2019cwq,Apolinario:2014csa,Mehtar-Tani:2019ygg,Andres:2020vxs}. When no approximation in the number of scatterings is employed, the evolution of $\hat{q}$ as a function of the temperature was recently shown to become smooth (see green squares in Fig.~\ref{fig:qhatFeal}). Within such prescription, the $\hat{q}$ parameter depends only on the temperature of the medium that is created.

A different approach to solve JET \textit{puzzle} has recently been put forward in \cite{Kumar:2019uvu}. In this work, it is argued that the exchanged momentum between the incoming hard parton and the medium varies over a range of scales. The $\hat{q}$ parameter can thus be expressed in terms of a QGP parton distribution function (QGP-PDF) that is used to calculate energy loss in the higher-twist approach. Within such framework, the $\hat{q}$ dependence on the scale of the jet is responsible for the different $\hat{q}$ normalisations obtained from fits to RHIC and LHC data. 

Progress has also been made in the heavy-quark description. As with the JET collaboration, six transport models for charm mesons were used to extract the drag and transport coefficient from experimental data on charm D-meson suppression in central Pb+Pb collisions at LHC. The obtained results are shown in~\cite{Cao:2018ews}, and further discussed in~\cite{Cao}.

\subsection{Thermalisation}

\begin{figure}[h]
\centering
 \begin{subfigure}[c]{0.32\textwidth}
 \includegraphics[width=\textwidth]{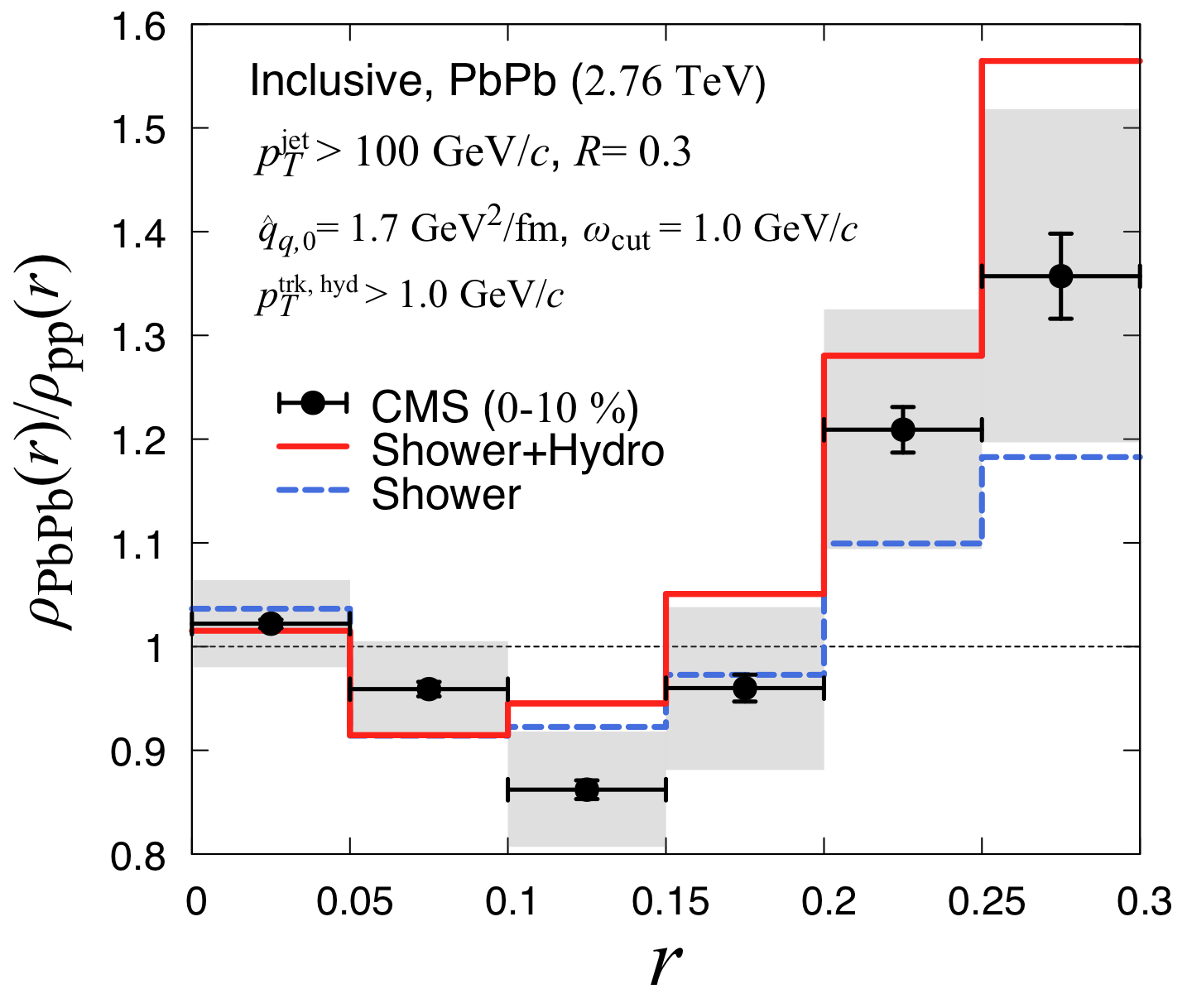}
 \caption{Jet Coupled Fluid~\cite{Tachibana:2017syd}.}
 \label{fig:jetProfileCoupled}
 \end{subfigure}
 \begin{subfigure}[c]{0.33\textwidth}
 \includegraphics[width=\textwidth]{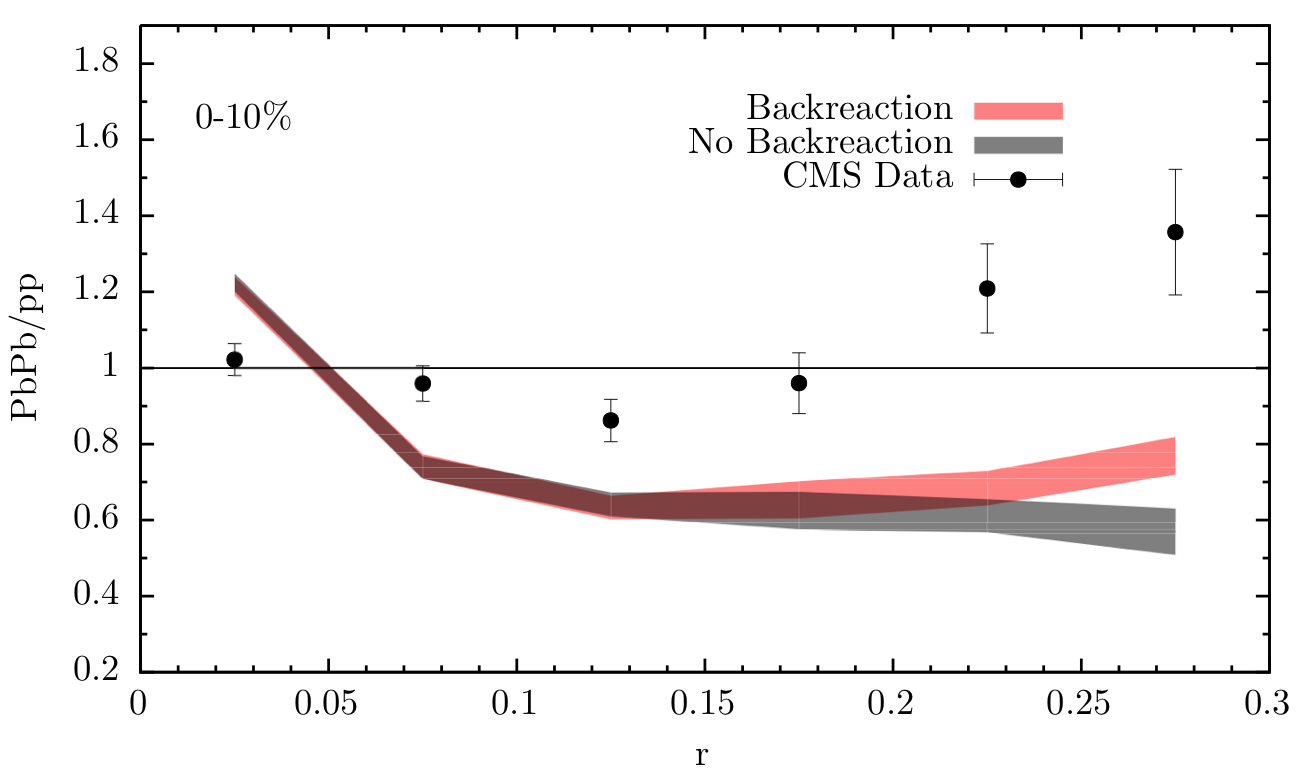}
 \caption{Hybrid Strong/Weak coupling approach~\cite{Casalderrey-Solana:2016jvj}.}
 \label{fig:jetProfileHybrid}
 \end{subfigure}
 \begin{subfigure}[c]{0.32\textwidth}
 \includegraphics[width=\textwidth]{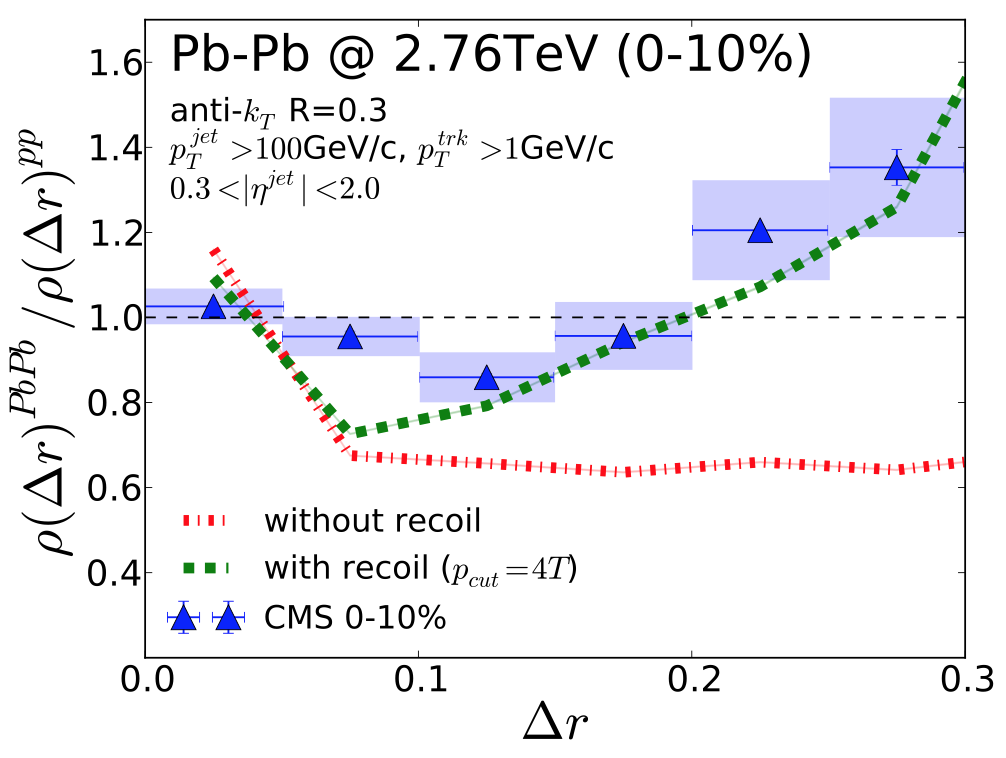}
 \caption{MARTINI~\cite{Park:2018acg}.}
 \label{fig:jetProfileMartini}
 \end{subfigure}
 \caption{Ratio of the jet radial profile from central Pb+Pb collisions to p+p collisions, at $\sqrt{s} = 2.76$~TeV, as measured by CMS, and comparison with 3 different jet quenching models.}
 \label{fig:jetProfile}
\end{figure}

Another important property to our understanding of the created hot and dense medium is its transparency to the passage of a high momentum particle. When this particle interacts elastically with the QGP, it will deposit and share its energy among the QGP constituents. The determination of how fast the jet energy is propagated and thermalised with the rest of the QGP (medium response) would allow us to understand its inner dynamics. The difficulty in assessing such property is to know how much this effect contributes to the final jet observables, and if so, what would be the measurement that is most sensitive to it. So far, the jet radial profile seems to be most susceptible to this contribution. However, in this conference, jet hadro-chemistry~\cite{ChenJetI} was also pointed as an additional observable to constrain the medium response. As an example, in Fig.~\ref{fig:jetProfile}, it is shown the CMS results on jet radial profile for Pb+Pb $[0-10]\%$ central collisions at 2.76~TeV ratio to p+p collisions, for 3 different jet quenching descriptions: a semi-analytical approach~\cite{Tachibana:2017syd} (Fig.~\ref{fig:jetProfileCoupled}), a hybrid strong/weak coupling model~\cite{Casalderrey-Solana:2016jvj} (Fig.~\ref{fig:jetProfileHybrid}), and a perturbative jet quenching Monte Carlo generator~\cite{Park:2018acg}(Fig.~\ref{fig:jetProfileMartini}).

\begin{figure}[h]
\centering
 \includegraphics[width=0.65\textwidth]{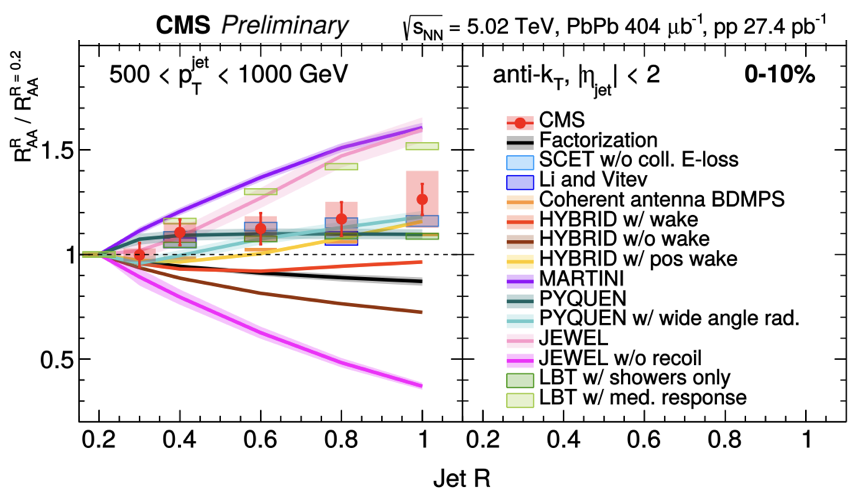}
 \caption{Jet nuclear suppression for Pb+Pb collisions at $\sqrt{s_{NN}} = 5.02$~TeV measured as the ratio for reconstructed anti-$k_T$ jets with an $R = 0.3; 0.4; 0.6; 0.8; 1.0$ with respect to reconstructed anti-$k_T$ jets with $R = 0.2$. The red dots with boxes represent the CMS data for $[0-10]\%$ central collisions and jets with a transverse momentum of$500 < p_T^{jet} < 100$~GeV. The remaining lines and boxes show a comparison for several models, some of them with and without medium response. Figure taken from~\cite{TaylorJetIII}.}
 \label{fig:cms}
\end{figure}

There are several uncertainties in the magnitude of this effect as it significantly varies between approaches. However, it seems to be a necessary feature to describe the excess of particles at large angles. In this conference, CMS collaboration showed some preliminary results on the jet suppression for very large-radius jets, up to $R = 1.0$ (see Fig.~\ref{fig:cms}, taken from~\cite{TaylorJetIII}). The ratio of the jet suppression with respect to small radius jets ($R = 0.2$) show a slight increase with the jet radius. This feature is not seen in the models that do need a significant amount of medium response to describe the jet radial profile. On the other hand, models that lack this contribution seem to capture relatively well this behaviour. Further comparisons of this new observable with the jet radial profile will put severe constraints on the jet-induced component allowing us to understand how exactly is the energy thermalised with the rest of the medium and/or what are the features of the parton shower itself that drive the behaviour of the jet suppression and jet radial profile.

\subsection{QGP Constitution}

\begin{figure}[h]
\centering
 \begin{subfigure}[c]{0.4\textwidth}
 \includegraphics[width=\textwidth]{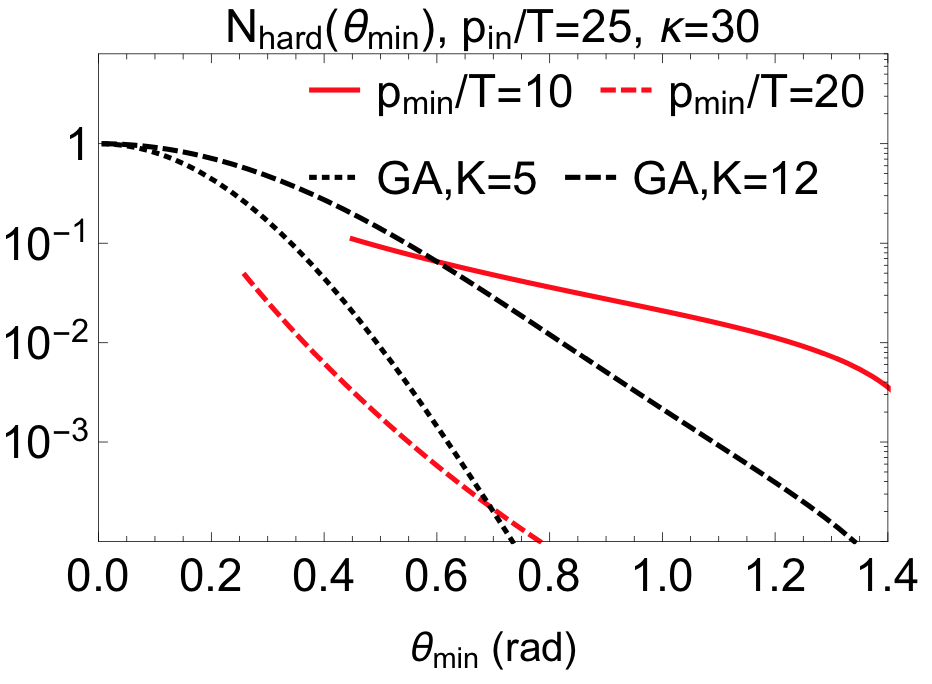}
 \end{subfigure}
 \hspace{5mm}
 \begin{subfigure}[c]{0.4\textwidth}
 \includegraphics[width=\textwidth]{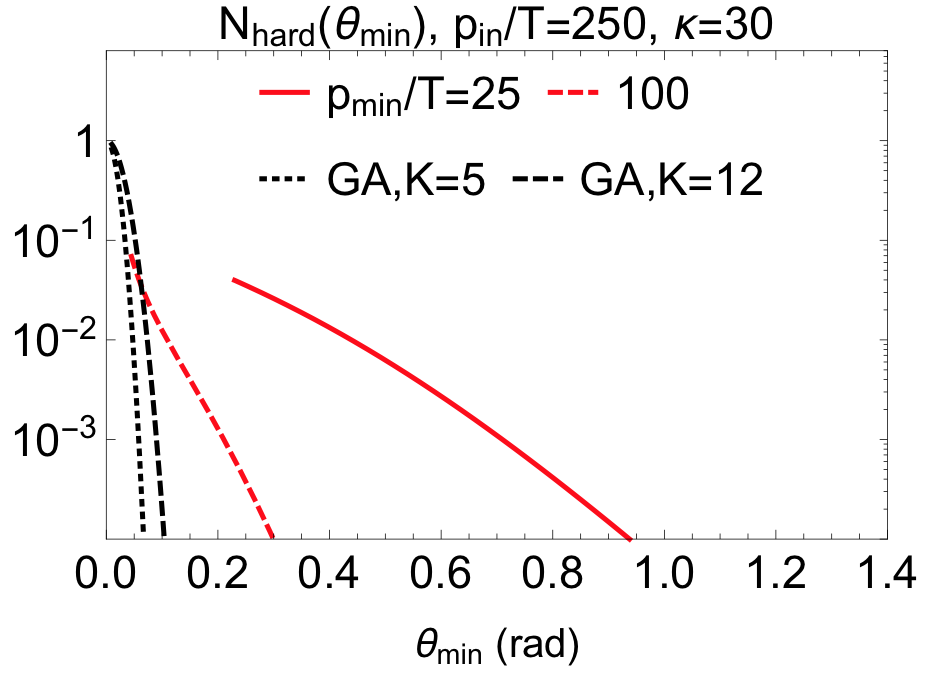}
 \end{subfigure}
 \caption{Number of outgoing particles with $\theta > \theta_{min}$ per incident gluon with $p_{in}/T = 25$ (left) and $p_{in}/T = 100$ (right). The solid red curve correspond to an outgoing particle with $p > p_{min} = 10 (25) T$ and the dashed curve for $p > p_{min} = 20 (100) T$ in the left (right) figure, that is deflected due to a single hard scattering. At black, it is shown the estimates using multiple soft scattering, valid only for small angles. Fig. taken from~\cite{DEramo:2018eoy}.}
 \label{fig:jetQuasi}
\end{figure}

In our pursuit of describing how the parton shower and high momentum particles are modified when propagating through a hot and dense medium, we need to understand how to characterise the intrinsic constitution of the QGP. Choices vary from modelling it as a collection of quasi-particles or by a strongly coupled fluid, without correspondence to hard thermal particles. To understand what is the QGP intrinsic constitution, is it possible to devise strategies to exclude one or the other scenario~\cite{DEramo:2012uzl,DEramo:2018eoy}. One of the possibilities is to find rare large-angle deflections of partons resulting from the interaction of the parton shower fragments with QGP particles. This type of measurements would put limits to the shortest distance between weakly coupled scatterers within the medium that is produced in heavy-ion collisions. In~\cite{DEramo:2018eoy}, parameterising the QGP as a \emph{brick} of quasi-particles at a given temperature, it was estimated the probability of finding a parton with a minimum momentum above the medium temperature, $p_{min}$, above a given angle, $\theta_{min}$, measured with respect to the initial incoming parton direction. The results for the number of outgoing hard partons at $\theta > \theta_{min}$ per incident gluon is shown in Fig.~\ref{fig:jetQuasi}. It is assumed an incoming gluon with an energy of $p_{in} = 25 T$ (Fig.\ref{fig:jetQuasi} left) and $p_{in} = 100 T$ (Fig.\ref{fig:jetQuasi} right). At red, it is shown the contribution due to single hard scattering, while the black lines correspond to multiple soft scattering, valid at small deflection angles. The larger the momentum required for the outgoing particles, the less probable is for the particle to be deflected at large angles. The signature requires high statistics, but with the future LHC upgrades, this proposal will be critical for determining the QGP constitution.

\subsection{QGP Resolution}

In~\cite{MehtarTani:2010ma,CasalderreySolana:2011rz}, it was shown that in the presence of medium-induced radiation, the phase space for the subsequent radiation during the parton shower evolution opens up. Coherent vacuum-like emissions dominantly follow the angular ordering behaviour, while medium-induced radiation is preferably emitted at large angles (anti-angular ordering). The interplay between such coherent/decoherent emissions should be relevant for the shower development. So far, the only Monte Carlo implementation of this effect is in the hybrid strong/weak approach~\cite{Hulcher:2017cpt}. In this model, energy loss is applied to a fully developed Pythia shower. To account for coherence effects, at each splitting, a coherence time is assigned to both daughters, during which, both particles lose energy according to their parent parton flavour and energy if their transverse separation is smaller than the medium resolution scale. Nonetheless, it was shown that the jet fragmentation is almost unmodified concerning the fully decoherent case, an observation that is in contradiction with analytical results~\cite{Mehtar-Tani:2014yea}. 

\begin{figure}[h]
\centering
 \includegraphics[width=0.7\textwidth]{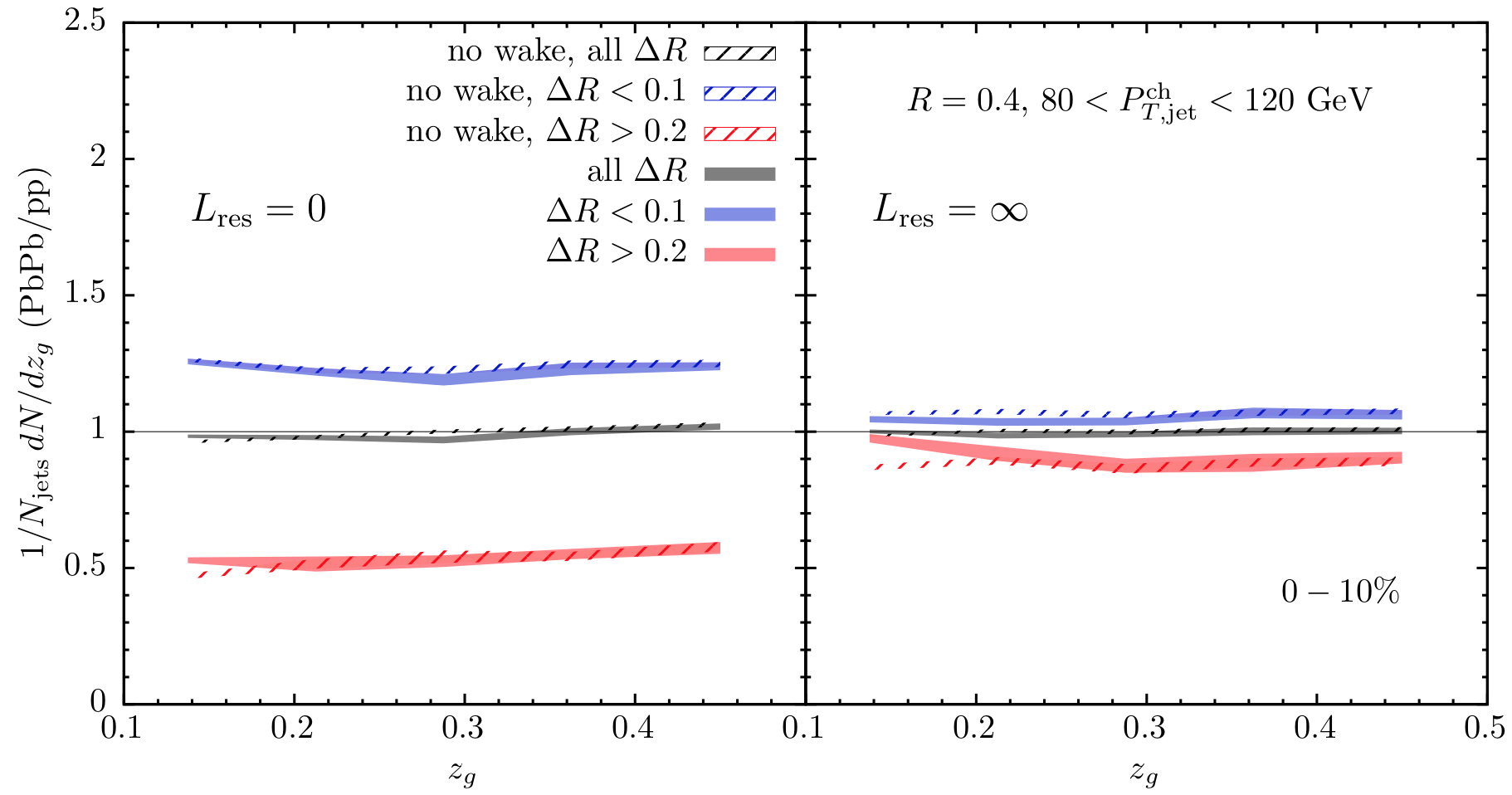}
 \caption{Ratio of the $z_g$ distribution for the most central PbPb collisions with $\sqrt{s} = 2.76$~TeV, obtained with the hybrid approach. In the left panel, the resolution scale $L_{res} = 0$, while on the right pane, $L_{res} = \infty$. The different colours represent different sub-jet separation in $\Delta R$. Figure taken from \cite{Casalderrey-Solana:2019ubu}.}
 \label{fig:HybridZg}
\end{figure}

Recently, it has been shown that the jet splitting function, $z_g$, can be a sensitive probe of the QGP resolution scale. In Fig.~\ref{fig:HybridZg}, using the hybrid Monte Carlo event generator, it is shown the ratio of the $z_g$ distribution (not self-normalised) for a jet with $R =0.4$ concerning the model reference. The grey bands correspond to all pair of sub-jets that are found when using the SoftDrop procedure, while at blue (salmon), only the pair of subjets that has a $\Delta R < 0.1$ ($\Delta R > 0.2$). When it is assumed that the shower development is completely decoherent ($L = 0$, left panel of Fig.~\ref{fig:HybridZg}), and therefore energy loss is applied to all shower partons, the resulting $z_g$ ratio shows a clear separation between the collinear jets (blue) and wide angle jets (salmon). Naturally, moving to the opposite limit (coherent shower, $L = \infty$, right panel of Fig.~\ref{fig:HybridZg}), these two populations will lose the same amount of energy, yield a similar $z_g$ distribution to vacuum. 

Interestingly, a comparison with ALICE (folded) data seems to disfavour the case where the jet is unresolved/coherent by the medium and instead seems to indicate the fully decoherent picture~\cite{PablosJetIII}. Future unfolded measurements will allow us to identify the QGP resolution scale, as well as going to higher jet transverse momentum.

\subsection{Small Systems}

Alongside the QGP properties, it is necessary to understand the required conditions to produce such state of hot and dense matter. In particular, when looking to p+A systems, where, so far, energy loss effects are absent. Nonetheless, a puzzling non-zero elliptic flow, $v_2$, component is quite visible in the results shown by ATLAS, for p+Pb collisions at $\sqrt{s} = 8.16$~TeV~\cite{SS2}. While this puzzle is yet to be solved, the perspectives for using lighter systems in the future HL/HE-LHC can enhance the heavy-ion programme with a vast number of rare probes that can potentially answer to this puzzling observation~\cite{Citron:2018lsq}.

\section{Into the Future: a personal and biased view}
\label{sec:3}

So far, we were able to measure and propose novel analysis that will, in a near future, lead to a deeper understanding of the QGP properties. The natural step is to bring our qualitative insight into a more quantitative analysis and extract the QGP properties. For that, not only we need higher statistics to reduce our statistical uncertainties of rare probes, but we also need to increase features independence. A natural example are boson+jet or boson+hadron observables that, while relatively rare, provide a natural and calibrated probe for energy loss studies, without the need for comparison with p+p collisions. Along this line, there has been the proposal of using fully reconstructed sub-jet observables that are, by construction, more independent of medium response effects~\cite{Apolinario:2017qay}. In this conference, some preliminary results on the use of fully reconstructed sub-jets were presented by STAR~\cite{RaghavJetII}. Not only they can be used to constrain parton energy loss, as they allow to create samples of collinear jets in both $p+p$ and $A+A$ collisions. Such class separation can potentiate an \textit{apples-to-apples} comparison, as well as to provide significant insight into the QGP resolution scale. As an alternative way to improve the matching of jets between $A+A$ and $p+p$, in~\cite{Brewer:2018dfs}, it was proposed to use the quantile ratio of the integrated cross-section for $A+A$ with respect to $p+p$ collisions. Because energy loss effects naturally bring $A+A$ jets to a different jet transverse momentum, the proposed ratio assures that the same fragmentation pattern can be compared with and without further interaction with the medium. The new large cone jets that were shown by both ATLAS~\cite{RybarJetIII} and CMS~\cite{TaylorJetIII} collaborations will also contribute to the feature independence mentioned above, in particular, to the energy loss fluctuations. When reconstructing small jet radius (from 0.4 to 0.5), there is a natural selection bias towards jets that did not interact much with the medium~\cite{Brewer:2017fqy,Milhano:2015mng} (jets that experienced larger than average energy loss fluctuations usually do not appear in the final sample as they lost too much energy). However, by increasing jet radius up to $R = 1$, the particles resulting from such positive fluctuations will be included in resulting final jet. The reduction of such selection bias will unlock more detailed studies on energy loss fluctuations. A quantitative assessment of the QGP characteristics also demands an accurate evaluation of the theoretical uncertainties. For such systematic model-data comparison, the JETSCAPE collaboration will provide the tool to understand each model features that are essential to describe the experimental observations. For more details on the achievements, we refer the reader for~\cite{SoltzJETSCAPE}.

A key feature that is currently being pursued along many different directions is the QGP time evolution. Such efforts are urgently needed as the medium is fast expanding (and so the QGP characteristics are highly time-dependent). Moreover, the puzzling observations on small systems are currently posing challenges to our understanding of the applicability of fluid behaviour, initially thought to be valid to extended strongly interacting matter. Along the possible exploration avenues, several works concentrated their attention into further constrain the uncertainties in the initialisation time of the QGP as a collective thermalised fluid. In particular, in~\cite{Andres:2019eus}, a combination of particle nuclear modification factor, $R_{AA}$, and $v_2$, can provide further constraints on initial anisotropy. Tops have also been proposed as time-delayed probes of the QGP, able to provide a detailed time-differential measurement~\cite{Apolinario:2017sob,Abada:2019lih}. First experimental observation of top-initiated jets was reported by the CMS collaboration~\cite{KrintirasJetIV}, yielding a $\sim 4 \sigma$ statistical significance. While limited by the current statistics, future measurements can provide limits on the QGP timescale. 

To extend the possibilities to perform a full QGP tomography is for sure among the community efforts. In~\cite{Mine}, there is a preliminary study that indicates how jets can be used to probe the different timescales of the QGP, in particular, how would be possible to identify the parton shower emissions that are parametrically emitted at the end of the medium (and thus unmodified with respect to vacuum shower), with respect to early emissions. By comparing those to the $p+p$ counterpart, it should be possible to withdraw meaningful information of the QGP characteristics and its fast evolution. Along this line of work, in~\cite{TakacsJT29}, it is shown how different parton shower orderings can affect Lund Planes and resulting jet multiplicity. Complementary with such Monte Carlo studies, ALICE~\cite{HavenerJetII} showed the first results on using Lund Planes to evaluate the kinematics and respective formation time of parton shower emissions, identified by different unclustering algorithms. These results seem to indicate a suppression of earlier (large angle) splittings and respective enhancement of late time (collinear) splittings. These studies will not only contribute to analyse the QGP time evolution, as proposed in~\cite{Mine}, but possible ways to use these results to unfold substructure variables in Pb+Pb collisions will also contribute for an accurate model-data comparison.

\section{Summary}
\label{sec:4}

Jets and single-particle measurements are complementary probes of the QGP. Both can be used for energy loss studies, transverse momentum broadening, and thermalisation of medium response. However, because jets are the result of the parton shower that propagated through the fast-expanding medium, they have the unique potential to reveal information about the QGP time structure and in-medium coherence properties. 

So far, we have achieved a vast number of developments that allowed to consolidate our understanding of in-medium parton shower, with subsequent extraction of the QGP properties. However, there are still several open questions, that will for sure pose us a variety of interesting challenged ahead. I trust the next years in this field to deliver fascinating results! \hfill \break

\textbf{Acknowledgments:} The author would like to thank N. Armesto, E. Ferreiro, A. Kurkela, Y-J. Lee, G. Milhano and C. A Salgado for their useful comments and suggestions. This work was supported by Funda\c{c}\~{a}o para a Ci\^{e}ncia e Tecnologia (FCT - Portugal) under projects DL57/2016/CP1345/ CT0004 and CERN/FIS-PAR/0022/2017 and European Research Council Grant under reference ERC-2018-ADG-835105 YoctoLHC.





\bibliographystyle{elsarticle-num}
\bibliography{Bibliography.bib}

\begin{thebibliography}{10}
\expandafter\ifx\csname url\endcsname\relax
  \def\url#1{\texttt{#1}}\fi
\expandafter\ifx\csname urlprefix\endcsname\relax\def\urlprefix{URL }\fi
\expandafter\ifx\csname href\endcsname\relax
  \def\href#1#2{#2} \def\path#1{#1}\fi

\bibitem{Luo}
T.~Luo, {Jet quenching and medium response}, These proceedings.

\bibitem{Tywoniuk}
K.~Tywoniuk, {Parton propagation and energy loss: new theoretical progress},
  These proceedings.

\bibitem{Cao}
S.~Cao, {Heavy quark transport: a theoretical overview}, These proceedings.

\bibitem{Wang}
J.~Wang, {Heavy Quark production and energy loss: experiments}, These
  proceedings.

\bibitem{Chen}
Y.~Chen, {Jet substructure and parton splitting: an experimental overview},
  These proceedings.

\bibitem{Trzeciak}
B.~Trzeciak, {Quenching of heavy and light flavor jets: experimental overview},
  These proceedings.

\bibitem{Moore:2004tg}
G.~D. Moore, D.~Teaney, {How much do heavy quarks thermalize in a heavy ion
  collision?}, Phys. Rev. C71 (2005) 064904.
\newblock \href {http://arxiv.org/abs/hep-ph/0412346}
  {\path{arXiv:hep-ph/0412346}}, \href
  {http://dx.doi.org/10.1103/PhysRevC.71.064904}
  {\path{doi:10.1103/PhysRevC.71.064904}}.

\bibitem{Qin:2009gw}
G.-Y. Qin, A.~Majumder, {A pQCD-based description of heavy and light flavor jet
  quenching}, Phys. Rev. Lett. 105 (2010) 262301.
\newblock \href {http://arxiv.org/abs/0910.3016} {\path{arXiv:0910.3016}},
  \href {http://dx.doi.org/10.1103/PhysRevLett.105.262301}
  {\path{doi:10.1103/PhysRevLett.105.262301}}.

\bibitem{Burke:2013yra}
K.~M. Burke, et~al., {Extracting the jet transport coefficient from jet
  quenching in high-energy heavy-ion collisions}, Phys. Rev. C90~(1) (2014)
  014909.
\newblock \href {http://arxiv.org/abs/1312.5003} {\path{arXiv:1312.5003}},
  \href {http://dx.doi.org/10.1103/PhysRevC.90.014909}
  {\path{doi:10.1103/PhysRevC.90.014909}}.

\bibitem{Feal:2019xfl}
X.~Feal, C.~A. Salgado, R.~A. Vazquez, {Jet quenching tests of the QCD Equation
  of State}\href {http://arxiv.org/abs/1911.01309} {\path{arXiv:1911.01309}}.

\bibitem{Andres:2016iys}
C.~Andrés, N.~Armesto, M.~Luzum, C.~A. Salgado, P.~Zurita, {Energy versus
  centrality dependence of the jet quenching parameter $\hat{q}$ at RHIC and
  LHC: a new puzzle?}, Eur. Phys. J. C76~(9) (2016) 475.
\newblock \href {http://arxiv.org/abs/1606.04837} {\path{arXiv:1606.04837}},
  \href {http://dx.doi.org/10.1140/epjc/s10052-016-4320-5}
  {\path{doi:10.1140/epjc/s10052-016-4320-5}}.

\bibitem{Sievert:2019cwq}
M.~D. Sievert, I.~Vitev, B.~Yoon, {A complete set of in-medium splitting
  functions to any order in opacity}, Phys. Lett. B795 (2019) 502--510.
\newblock \href {http://arxiv.org/abs/1903.06170} {\path{arXiv:1903.06170}},
  \href {http://dx.doi.org/10.1016/j.physletb.2019.06.019}
  {\path{doi:10.1016/j.physletb.2019.06.019}}.

\bibitem{Apolinario:2014csa}
L.~Apolinário, N.~Armesto, J.~G. Milhano, C.~A. Salgado, {Medium-induced gluon
  radiation and colour decoherence beyond the soft approximation}, JHEP 02
  (2015) 119.
\newblock \href {http://arxiv.org/abs/1407.0599} {\path{arXiv:1407.0599}},
  \href {http://dx.doi.org/10.1007/JHEP02(2015)119}
  {\path{doi:10.1007/JHEP02(2015)119}}.

\bibitem{Mehtar-Tani:2019ygg}
Y.~Mehtar-Tani, K.~Tywoniuk, {Improved opacity expansion for medium-induced
  parton splitting }\href {http://arxiv.org/abs/1910.02032}
  {\path{arXiv:1910.02032}}.

\bibitem{Andres:2020vxs}
C.~Andres, L.~Apolin\'{a}rio, F.~Dominguez, {Medium-induced gluon radiation
  with full resummation of multiple scatterings for realistic parton-medium
  interactions }\href {http://arxiv.org/abs/2002.01517}
  {\path{arXiv:2002.01517}}.

\bibitem{Kumar:2019uvu}
A.~Kumar, A.~Majumder, C.~Shen, {Energy and scale dependence of $\hat{q}$ and
  the ``JET puzzle''}, Phys. Rev. C 101~(3) (2020) 034908.
\newblock \href {http://arxiv.org/abs/1909.03178} {\path{arXiv:1909.03178}},
  \href {http://dx.doi.org/10.1103/PhysRevC.101.034908}
  {\path{doi:10.1103/PhysRevC.101.034908}}.

\bibitem{Cao:2018ews}
S.~Cao, et~al., {Toward the determination of heavy-quark transport coefficients
  in quark-gluon plasma}, Phys. Rev. C99~(5) (2019) 054907.
\newblock \href {http://arxiv.org/abs/1809.07894} {\path{arXiv:1809.07894}},
  \href {http://dx.doi.org/10.1103/PhysRevC.99.054907}
  {\path{doi:10.1103/PhysRevC.99.054907}}.

\bibitem{ChenJetI}
C.~Wei, T.~Luo, S.~Cao, L.~Pang, X.~Wang, H.~M., {The medium modification of
  jet fragmentation function and baryon-to-meson ratio in jet}, These
  proceedings.

\bibitem{Tachibana:2017syd}
Y.~Tachibana, N.-B. Chang, G.-Y. Qin, {Full jet in quark-gluon plasma with
  hydrodynamic medium response}, Phys. Rev. C95~(4) (2017) 044909.
\newblock \href {http://arxiv.org/abs/1701.07951} {\path{arXiv:1701.07951}},
  \href {http://dx.doi.org/10.1103/PhysRevC.95.044909}
  {\path{doi:10.1103/PhysRevC.95.044909}}.

\bibitem{Casalderrey-Solana:2016jvj}
J.~Casalderrey-Solana, D.~Gulhan, G.~Milhano, D.~Pablos, K.~Rajagopal, {Angular
  Structure of Jet Quenching Within a Hybrid Strong/Weak Coupling Model}, JHEP
  03 (2017) 135.
\newblock \href {http://arxiv.org/abs/1609.05842} {\path{arXiv:1609.05842}},
  \href {http://dx.doi.org/10.1007/JHEP03(2017)135}
  {\path{doi:10.1007/JHEP03(2017)135}}.

\bibitem{Park:2018acg}
C.~Park, S.~Jeon, C.~Gale, {Jet modification with medium recoil in quark-gluon
  plasma}, Nucl. Phys. A982 (2019) 643--646.
\newblock \href {http://arxiv.org/abs/1807.06550} {\path{arXiv:1807.06550}},
  \href {http://dx.doi.org/10.1016/j.nuclphysa.2018.10.057}
  {\path{doi:10.1016/j.nuclphysa.2018.10.057}}.

\bibitem{TaylorJetIII}
M.~Taylor, {Mapping the redistribution of jet energy in PbPb collisions using
  jets with various radius parameters with CMS}, These proceedings.

\bibitem{DEramo:2012uzl}
F.~D'Eramo, M.~Lekaveckas, H.~Liu, K.~Rajagopal, {Momentum Broadening in Weakly
  Coupled Quark-Gluon Plasma (with a view to finding the quasiparticles within
  liquid quark-gluon plasma)}, JHEP 05 (2013) 031.
\newblock \href {http://arxiv.org/abs/1211.1922} {\path{arXiv:1211.1922}},
  \href {http://dx.doi.org/10.1007/JHEP05(2013)031}
  {\path{doi:10.1007/JHEP05(2013)031}}.

\bibitem{DEramo:2018eoy}
F.~D'Eramo, K.~Rajagopal, Y.~Yin, {Molière scattering in quark-gluon plasma:
  finding point-like scatterers in a liquid}, JHEP 01 (2019) 172.
\newblock \href {http://arxiv.org/abs/1808.03250} {\path{arXiv:1808.03250}},
  \href {http://dx.doi.org/10.1007/JHEP01(2019)172}
  {\path{doi:10.1007/JHEP01(2019)172}}.

\bibitem{MehtarTani:2010ma}
Y.~Mehtar-Tani, C.~A. Salgado, K.~Tywoniuk, {Anti-angular ordering of gluon
  radiation in QCD media}, Phys. Rev. Lett. 106 (2011) 122002.
\newblock \href {http://arxiv.org/abs/1009.2965} {\path{arXiv:1009.2965}},
  \href {http://dx.doi.org/10.1103/PhysRevLett.106.122002}
  {\path{doi:10.1103/PhysRevLett.106.122002}}.

\bibitem{CasalderreySolana:2011rz}
J.~Casalderrey-Solana, E.~Iancu, {Interference effects in medium-induced gluon
  radiation}, JHEP 08 (2011) 015.
\newblock \href {http://arxiv.org/abs/1105.1760} {\path{arXiv:1105.1760}},
  \href {http://dx.doi.org/10.1007/JHEP08(2011)015}
  {\path{doi:10.1007/JHEP08(2011)015}}.

\bibitem{Hulcher:2017cpt}
Z.~Hulcher, D.~Pablos, K.~Rajagopal, {Resolution Effects in the Hybrid
  Strong/Weak Coupling Model}, JHEP 03 (2018) 010.
\newblock \href {http://arxiv.org/abs/1707.05245} {\path{arXiv:1707.05245}},
  \href {http://dx.doi.org/10.1007/JHEP03(2018)010}
  {\path{doi:10.1007/JHEP03(2018)010}}.

\bibitem{Mehtar-Tani:2014yea}
Y.~Mehtar-Tani, K.~Tywoniuk, {Jet (de)coherence in Pb?Pb collisions at the
  LHC}, Phys. Lett. B744 (2015) 284--287.
\newblock \href {http://arxiv.org/abs/1401.8293} {\path{arXiv:1401.8293}},
  \href {http://dx.doi.org/10.1016/j.physletb.2015.03.041}
  {\path{doi:10.1016/j.physletb.2015.03.041}}.

\bibitem{Casalderrey-Solana:2019ubu}
J.~Casalderrey-Solana, G.~Milhano, D.~Pablos, K.~Rajagopal, {Modification of
  Jet Substructure in Heavy Ion Collisions as a Probe of the Resolution Length
  of Quark-Gluon Plasma}, JHEP 01 (2020) 044.
\newblock \href {http://arxiv.org/abs/1907.11248} {\path{arXiv:1907.11248}},
  \href {http://dx.doi.org/10.1007/JHEP01(2020)044}
  {\path{doi:10.1007/JHEP01(2020)044}}.

\bibitem{PablosJetIII}
J.~Casalderrey-Solana, G.~Milhano, D.~Pablos, K.~Rajagopal, {Modification of
  jet substructure in heavy ion collisions as a probe of the resolution length
  of quark-gluon plasma}, These proceedings.

\bibitem{SS2}
K.~Hill, {Collective behavior of high-$p_T$ particles in 8.16 TeV $p+Pb$
  collisions with the ATLAS detector}, These proceedings.

\bibitem{Citron:2018lsq}
Z.~Citron, et~al., {Report from Working Group 5}, CERN Yellow Rep. Monogr. 7
  (2019) 1159--1410.
\newblock \href {http://arxiv.org/abs/1812.06772} {\path{arXiv:1812.06772}},
  \href {http://dx.doi.org/10.23731/CYRM-2019-007.1159}
  {\path{doi:10.23731/CYRM-2019-007.1159}}.

\bibitem{Apolinario:2017qay}
L.~Apolin\'{a}rio, J.~G. Milhano, M.~Ploskon, X.~Zhang, {Novel subjet
  observables for jet quenching in heavy-ion collisions}, Eur. Phys. J. C78~(6)
  (2018) 529.
\newblock \href {http://arxiv.org/abs/1710.07607} {\path{arXiv:1710.07607}},
  \href {http://dx.doi.org/10.1140/epjc/s10052-018-5999-2}
  {\path{doi:10.1140/epjc/s10052-018-5999-2}}.

\bibitem{RaghavJetII}
E.~Elayavalli, {Constraining parton energy loss via angular and momentum based
  differential jet measurements at STAR}, These proceedings.

\bibitem{Brewer:2018dfs}
J.~Brewer, J.~G. Milhano, J.~Thaler, {Sorting out quenched jets}, Phys. Rev.
  Lett. 122~(22) (2019) 222301.
\newblock \href {http://arxiv.org/abs/1812.05111} {\path{arXiv:1812.05111}},
  \href {http://dx.doi.org/10.1103/PhysRevLett.122.222301}
  {\path{doi:10.1103/PhysRevLett.122.222301}}.

\bibitem{RybarJetIII}
M.~Rybar, {Measurement of jet structure and substructure in heavy ion
  collisions with ATLAS}, These proceedings.

\bibitem{Brewer:2017fqy}
J.~Brewer, K.~Rajagopal, A.~Sadofyev, W.~Van Der~Schee, {Evolution of the Mean
  Jet Shape and Dijet Asymmetry Distribution of an Ensemble of Holographic Jets
  in Strongly Coupled Plasma}, JHEP 02 (2018) 015.
\newblock \href {http://arxiv.org/abs/1710.03237} {\path{arXiv:1710.03237}},
  \href {http://dx.doi.org/10.1007/JHEP02(2018)015}
  {\path{doi:10.1007/JHEP02(2018)015}}.

\bibitem{Milhano:2015mng}
J.~G. Milhano, K.~C. Zapp, {Origins of the di-jet asymmetry in heavy ion
  collisions}, Eur. Phys. J. C76~(5) (2016) 288.
\newblock \href {http://arxiv.org/abs/1512.08107} {\path{arXiv:1512.08107}},
  \href {http://dx.doi.org/10.1140/epjc/s10052-016-4130-9}
  {\path{doi:10.1140/epjc/s10052-016-4130-9}}.

\bibitem{SoltzJETSCAPE}
R.~Soltz, {A Comprehensive MC framework for jet quenching}, These proceedings.

\bibitem{Andres:2019eus}
C.~Andres, N.~Armesto, H.~Niemi, R.~Paatelainen, C.~A. Salgado, {Jet quenching
  as a probe of the initial stages in heavy-ion collisions}\href
  {http://arxiv.org/abs/1902.03231} {\path{arXiv:1902.03231}}.

\bibitem{Apolinario:2017sob}
L.~Apolin\'{a}rio, J.~G. Milhano, G.~P. Salam, C.~A. Salgado, {Probing the time
  structure of the quark-gluon plasma with top quarks}, Phys. Rev. Lett.
  120~(23) (2018) 232301.
\newblock \href {http://arxiv.org/abs/1711.03105} {\path{arXiv:1711.03105}},
  \href {http://dx.doi.org/10.1103/PhysRevLett.120.232301}
  {\path{doi:10.1103/PhysRevLett.120.232301}}.

\bibitem{Abada:2019lih}
A.~Abada, et~al., {FCC Physics Opportunities}, Eur. Phys. J. C79~(6) (2019)
  474.
\newblock \href {http://dx.doi.org/10.1140/epjc/s10052-019-6904-3}
  {\path{doi:10.1140/epjc/s10052-019-6904-3}}.

\bibitem{KrintirasJetIV}
F.~Krintiras, {Observation of top quark pair production in nucleus-nucleus
  collisions with the CMS detector}, These proceedings.

\bibitem{Mine}
L.~Apolin~\'{a}rio, {Time evolution of a medium-modified jet}, in: {2019
  European Physical Society Conference on High Energy Physics (EPS-HEP2019)
  Ghent, Belgium, July 10-17, 2019}, 2020.
\newblock \href {http://arxiv.org/abs/2001.06170} {\path{arXiv:2001.06170}}.

\bibitem{TakacsJT29}
A.~Takács, D.~Pablos, K.~Tywoniuk, {Can jet quenching constrain the evolution
  history of parton showers?}, These proceedings.

\bibitem{HavenerJetII}
L.~Havener, {Exploring the phase space of jet splittings in Pb-Pb and pp
  collisions at $\sqrt{s_{NN}}= 5.02$ TeV in ALICE}, These proceedings.

\end{thebibliography}




\end{document}